\documentclass[aps,prl,twocolumn,superscriptaddress]{revtex4-2}
\usepackage{graphicx}
\usepackage{amsmath,mathtools,amssymb}
\usepackage{hyperref}
\usepackage{float}

\begin{document}

\title{Probing Remote Nuclear Magnetic Moments in hBN with $V_B^-$ Electron Spin}

\author{G.V. Mamin}
\affiliation{Institute of Physics, Kazan Federal University, Kremlyovskaya 18, Kazan 420008, Russia}

\author{E.V. Dmitrieva}
\affiliation{Institute of Physics, Kazan Federal University, Kremlyovskaya 18, Kazan 420008, Russia}

\author{F.F. Murzakhanov}
\affiliation{Institute of Physics, Kazan Federal University, Kremlyovskaya 18, Kazan 420008, Russia}

\author{I.N. Gracheva}
\affiliation{Institute of Physics, Kazan Federal University, Kremlyovskaya 18, Kazan 420008, Russia}

\author{E.N. Mokhov}
\affiliation{Ioffe Institute, Polytekhnicheskaya, 26, St. Petersburg 194021, Russia}

\author{I.I. Vlasov}
\affiliation{Prokhorov General Physics Institute of the Russian Academy of Sciences, 119991 Moscow, Russia}

\author{M.R. Gafurov}
\affiliation{Institute of Physics, Kazan Federal University, Kremlyovskaya 18, Kazan 420008, Russia}

\author{U. Gerstmann}
\affiliation{Lehrstuhl f{\"u}r Theoretische Materialphysik, Universit{\"a}t Paderborn, 33098 Paderborn, Germany}

\author{V.A. Soltamov}
\email{Corresponding author.\\
victrosoltamov@gmail.com}
\affiliation{Ioffe Institute, Polytekhnicheskaya, 26, St. Petersburg 194021, Russia}

\email{victrosoltamov@gmail.com}

\date{\today}

\begin{abstract}
Since the initial discovery of optically addressable spins of the negatively charged boron vacancy defect ($V_B^-$) in hexagonal boron nitride (hBN), substantial progress has been made, enabling promising applications in quantum sensing, information processing, and simulations. A deep understanding of the $V_B^-$ (electron): hBN (nuclear) spin systems is crucial for realizing these potentials. In this article, we employ Electron Nuclear Double Resonance (ENDOR) to demonstrate the sensing of distant nuclear spins via the $V_B^-$ electron spin. We identify the nature and localization of the probed nuclear magnetic moments as $^{14}$N spins localized $\approx$~0.4~nm away from the vacancy and resolve the energies of the corresponding interactions. Density Functional Theory (DFT) calculations further confirm these findings, providing a detailed description of the interactions between the $V_B^-$ electron spin and surrounding nitrogen atoms in different shells. The results establish the $V_B^-$ electron spin as a promising tool for developing novel van der Waals material-based nuclear magnetic resonance (NMR) probes, advancing the understanding of spin physics in hBN, and unlocking its potential to study distant nuclear spin interactions in the host.
\end{abstract}

\maketitle

Hexagonal boron nitride (hBN), a prominent representative of van der Waals (vdW) materials, possesses exceptional potential as a platform for quantum technologies utilizing solid-state spin defects \cite{ref1, ref2, ref3}. Its ultra-wide bandgap ($E_g \approx 6$ eV) \cite{ref4}, exceeding that of diamond ($E_g \approx 5.5$ eV), enables the incorporation of a diverse range of optically active defects emitting across a broad spectral range, from deep ultraviolet (UV) to near-infrared (NIR), with the ability to study individual defects \cite{ref5, ref6, ref7}. The structural properties of hBN, characterized by honeycomb planes formed of strongly sp$^2$-hybridized boron and nitrogen atoms bound by weak vdW forces, enable precise nanoscale engineering down to the two-dimensional (2D) monolayer level \cite{ref8}. This, coupled with its exceptional thermal and chemical stability, positions hBN as a versatile platform for a wide range of advanced materials applications, including the synthesis of vdW heterostructures and the development of nanoelectronics and optoelectronic devices \cite{ref9, ref10}.

\begin{figure}
    \centering
    \includegraphics[width=\linewidth]{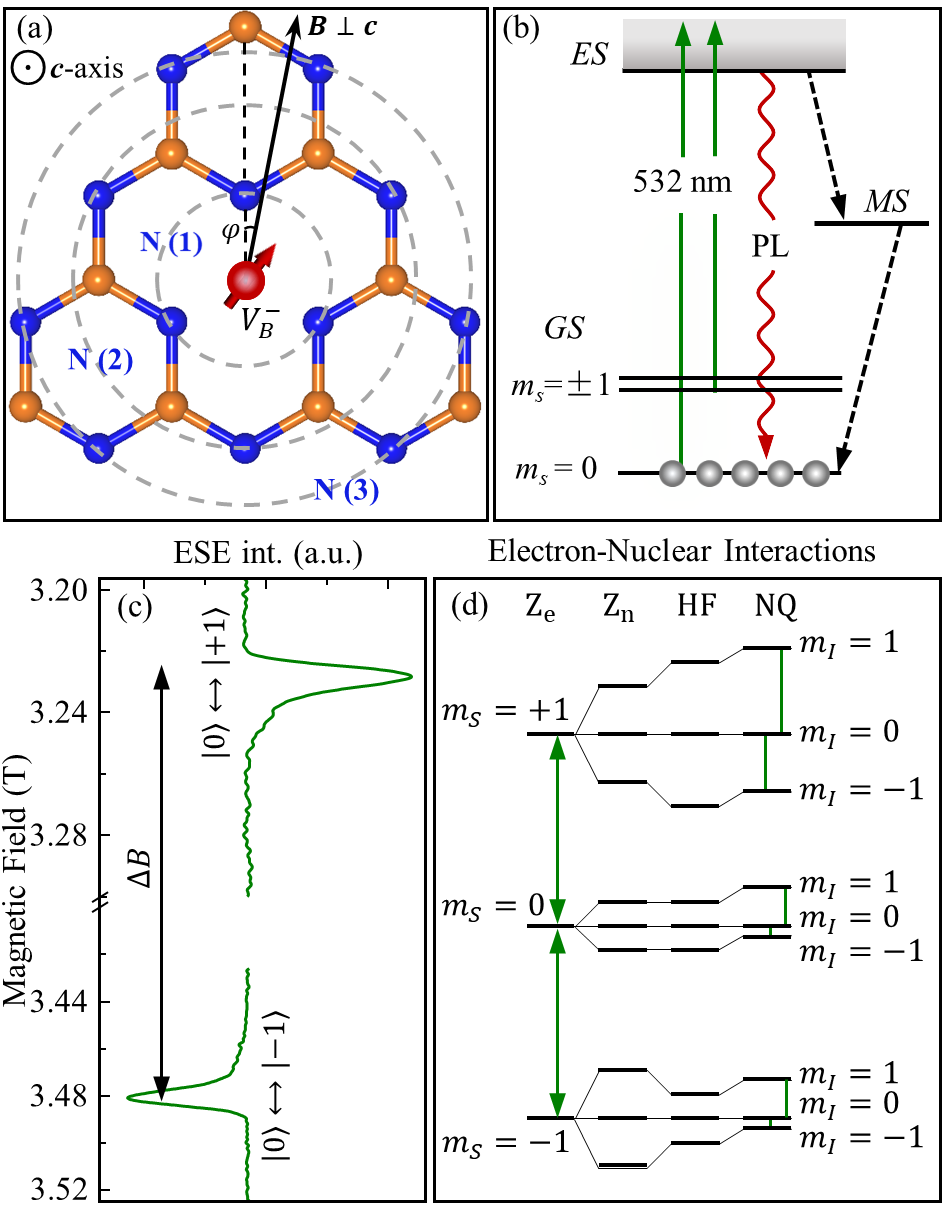}
    \caption{(a) hBN layer with $V_B^-$ defect (ball with arrow). The hexagonal $c$-axis is perpendicular to the figure plane. Nitrogen atoms groups, N(1), N(2), N(3), are shown with gray circles. N(1) and N(2) each contain 3 nitrogen atoms; N(3) has 6 atoms. The in-plane orientation of the $B$ field vector ($B\perp c$), making an angle $\varphi$ with the direction of the nitrogen dangling bond (denoted by the dashed line), is represented by a solid black arrow. (b) Diagram of $V_B^-$ ground state spin $m_S = 0$ initialization via spin-dependent recombination (dashed lines) under $\lambda=532$ nm excitation, with radiative decay (PL) shown in purple. (c) $V_B^-$ ESR spectrum at $B \parallel c$ under $\lambda=532$ nm excitation. ESR transitions ($\Delta m_S = \pm1$), separated by $\Delta B$, are indicated. (d) Energy levels of $V_B^-$ electron spin interacting with $I = 1$ $^{14}$N nuclei, including Zeeman (Ze), zero-field splitting (ZFS), hyperfine (HF), nuclear quadrupole (NQ), and nuclear Zeeman (Zn) interactions. Levels are labelled with $m_S, m_I$ quantum numbers. The allowed ESR and NMR transitions are shown with green bars.}
    \label{fig1}
\end{figure}

The discovery of optically addressable triplet ($S = 1$) spin states of negatively charged boron vacancy ($V_B^-$) defects in hBN (Fig.~\ref{fig1}a,b) \cite{ref1, ref11}, along with the ability to read out these spins using optically detected magnetic resonance (ODMR), has driven significant research into the $V_B^-$ spin-optical properties and their applications in quantum technologies. This line of investigation has also led to the identification of other defect systems in hBN, such as carbon-related defects, which also enable ODMR detection \cite{ref2, ref12, ref13, ref14}. These defect systems in hBN pave the way for applications in sensing \cite{ref15, ref16, ref17, ref18, ref19} and computation \cite{ref7, ref13, ref14, ref16, ref20, ref21, ref22, ref23}, while also marking the first steps toward developing quantum simulators based on the inherent nuclear spins of hBN \cite{ref21, ref22, ref23}. Interactions between the electron spin of defects and surrounding nuclear magnetic moments, mediated by hyperfine (HF) and nuclear quadrupole (NQ) interactions, play a central role in all aspects mentioned above. This is particularly significant since the hBN lattice exclusively comprises chemical elements with nonzero nuclear spins: 99.9\% $^{14}$N ($I = 1$), 19.9\% $^{10}$B ($I = 3$), and 80.1\% $^{11}$B ($I = 3/2$). To date, microwave spectroscopy has resolved and studied these interactions in detail for the three nearest nitrogen atoms around the boron vacancy (N(1) group in Fig.~\ref{fig1}a), yielding results consistent with DFT calculations \cite{ref1, ref24, ref25, ref26}. Also, HF interactions originating from $^{13}$C atoms intrinsically forming the carbon-related defect have been investigated in Ref.~\cite{ref14} using a combination of Electron Nuclear Double Resonance (ENDOR) and ODMR spectroscopy.

In this study, using high-frequency (94 GHz) ENDOR, we probe remote $^{14}$N magnetic moments, through the $V_B^-$ electron spin. Analysis of the ENDOR spectra recorded for distinct magnetic field orientations enables us to assign the probed nitrogen atoms to the third nitrogen shell, N(3). The variation in ENDOR frequencies across different orientations was then used to extract the corresponding HF and NQ tensors, with both the values and sources of these interactions confirmed by DFT calculations. This approach establishes the foundation for implementing the $V_B^-$ defect as a nuclear magnetic resonance (NMR) probe, identifying distant $^{14}$N spins for electron-nuclear register, and contributing critical insights into $V_B^-$ spin-Hamiltonian.

Commercial hBN single crystal (hq Graphene) irradiated at room temperature with 2 MeV electrons to a dose of $6 \times 10^{18}$ cm$^{-2}$ to generate $V_B^-$ defects \cite{ref27} was used. Experiments were conducted using a W-band (~94 GHz) Bruker Elexsys 680 ESR spectrometer. The electron spin echo (ESE)-detected ESR spectra were recorded using the standard Hahn-echo pulse sequence ($\pi/2$–$\tau$–$\pi$–$\tau$–ESE) with $\pi = 48$ ns and $\tau = 300$ ns. The $\pi/2$-pulse generates transverse magnetization, which decays over time $\tau$. The $\pi$-pulse refocuses spin packets, producing a spin-echo signal. To probe nuclear spins we use a standard Mims ENDOR pulse sequence $\pi/2$–$\tau$–$\pi/2$– $\pi_{RF}$– $\pi/2$–$\tau$–SSE \cite{ref28}. The first two $\pi/2$ mw pulses invert the electron spin population; the third $\pi/2$ pulse generates the stimulated electron spin echo (SSE) signal. Between the second and the third mw pulses, a radiofrequency $\pi_{RF}$ pulse is applied to invert the population of the nuclear spin sublevels, inducing NMR transitions. All experiments were conducted at a temperature $T = 25$ K.

The spectrum shown in Fig. 1(c) reveals two prominent ESR signals that serve as clear signatures of the $V_B^-$ defects, corresponding to magnetic dipole transitions within the triplet spin system. The signal separation, $\Delta B \approx 255$ mT, corresponds to the zero-field splitting value $D = \Delta B / (2g\mu_B) \approx 3.57$ GHz, where $g = 2.00$ is the Lande $g$-factor and $\mu_B$ is the Bohr magneton. These parameters confirm the spectroscopic characteristics of the $S = 1$ $V_B^-$ defect \cite{ref1, ref25}. The spectrum also shows that 532 nm excitation initializes the defect’s spin system into the $m_S = 0$ ground state via spin-selective intersystem crossing. This is evident from the phase reversal of the high-field ESR signal, which emits rather than absorbs microwave power.

We monitor the ESE signal intensity to study the $^{14}$N nuclear spins while driving NMR transitions between their spin sublevels. To analyze ESR and ENDOR data, the following spin-Hamiltonian is considered:

\begin{equation}
\setlength{\abovedisplayskip}{2pt}
\setlength{\belowdisplayskip}{2pt}
\vcenter{
\hbox{
$\begin{array}{rl}
H = & g \mu_B B_0 S_z + D(S_z^2 - S(S+1)/3) + \\[1ex]                                           
    & \hspace{-2.5em}\sum_i \big[ A_{zz} S_z I_{zi} + A_{xx} S_x I_{xi} + A_{yy} S_y I_{yi} \big] + \\[1ex]
    & \hspace{-2.5em}\sum_i \Big[ g_N \mu_N B_0 I_{zi} + P \big[(3 I_{zi}^2 - I(I+1)) + \eta (I_{xi}^2 - I_{yi}^2) \big]\Big]
\end{array}$}}
\end{equation}

Here $A_{xx}$, $A_{yy}$, $A_{zz}$ are the HF interaction values, $\mu_N$ and $g_N$ are the nuclear magneton and $^{14}$N nuclear $g$-factor, respectively. $P = \frac{C_q}{4I(2I-1)}$ describes the nuclear quadrupole interaction, where $C_q = eQ_N V_{zz}$ is the quadrupole coupling constant related to the interaction of the electric field gradient $V_{zz}$ with the nuclear electric quadrupole moment $Q_N$; $\eta = (P_{xx} - P_{yy})/P_{zz}$ quantifies the deviation from axial symmetry. $S_{x,y,z}$ and $I_{x,y,z}$ are the projections of the electron and nuclear spins, respectively.

The corresponding energy level diagram describing the $V_B^-$ electron spin interacting with $I = 1$ $^{14}$N nuclear spins localized in the 2D hBN plane, is presented in Fig. 1(d). It is important to note that, due to the high $D_{3h}$ symmetry of the $V_B^-$ defect, all in-plane nitrogen atoms within each group (N(1), N(2), and N(3)) are equivalent when the external magnetic field is oriented perpendicular to the (0001) plane ($B \parallel c$). Consequently, the frequencies of the NMR transitions at a given orientation are independent of the number of nitrogen atoms and can be treated as if they originate from the nuclear spin sublevels of a single $^{14}$N nucleus, as shown in Fig. 1(d). For each group of equivalent nitrogen atoms, six allowed NMR transitions ($\Delta m_I = \pm1$)are expected: HF and NQ interactions between the nuclear spin sublevels in the $m_S = +1$ and $m_S = -1$ states, and two pure quadrupole-induced transitions in the $m_S = 0$ state. These quadrupole-induced transitions are symmetrically positioned around the $^{14}$N Larmor frequency, which is about $\nu_L \approx 10$ MHz in the magnetic field of 3T. 
\begin{figure}
    \centering
    \includegraphics[width=\linewidth]{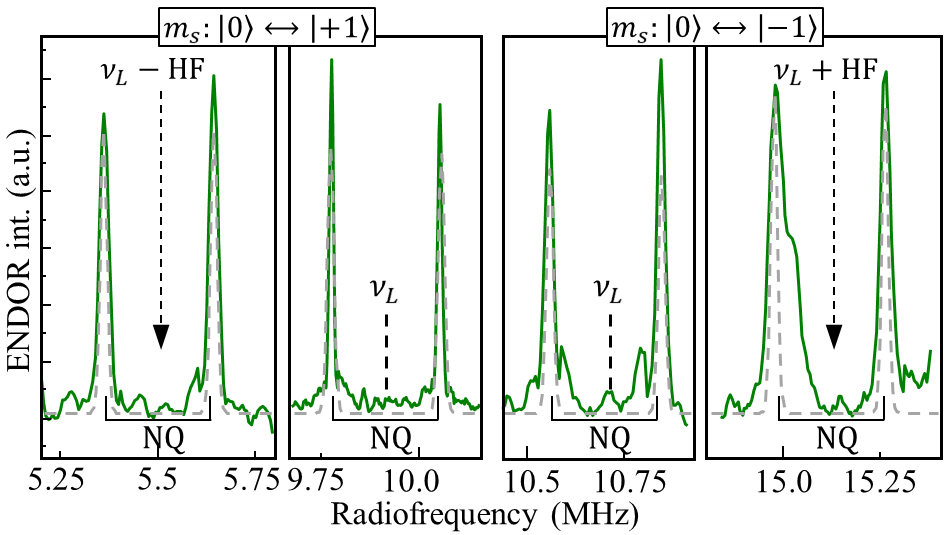} 
    \caption{NMR transitions measured by monitoring the SSE signal intensity of $V_B^-$ defect at the resonance magnetic fields $B = 3.2214$~T (corresponding to the $|0\rangle \leftrightarrow |+1\rangle$ ESR transition) and $B = 3.4765$~T ($|0\rangle \leftrightarrow |-1\rangle$ ESR transition). Experimental data are shown in green, and calculated data using Equation~(1) and EasySpin software~\cite{ref29} are shown in grey. The $^{14}$N Nuclear Larmor frequency ($\nu_L$) is indicated by a dashed vertical bar. Splitting of each pair of lines is induced by the NQ interaction as indicated by vertical bars. The HF interaction with $^{14}$N nuclei shifts each pair of quadrupole-split lines from $\nu_L$ by a constant value, as indicated by dashed arrows at $\nu_L\pm$HF.}
    \label{fig:ENDOR_spectrum}
\end{figure}
This exact pattern is observed in the ENDOR spectra measured for both ESR transitions ($|0\rangle \leftrightarrow |+1\rangle$ and $|0\rangle \leftrightarrow |-1\rangle$) at the $B\parallel c$ orientation, as shown in Fig.~2, unambiguously indicating the interactions of the $V_B^-$ electron spin with the $^{14}$N nuclear spins. The extracted HF and NQ splittings from the spectrum fitting using Eq.~(1) are approximately 4.41 MHz and 0.37 MHz, respectively. These values are notably smaller than those previously reported for $V_B^-$ interactions with the first nearest N(1) neighbors nitrogen spins, where the HF and NQ have been determined to be about 45 MHz and 1.5 MHz, respectively~\cite{ref1,ref24}.
To determine the values of the corresponding HF and NQ tensors as well as to assign the nitrogen atoms to specific shells relative to $V_B^-$, we measure the NMR transition frequencies between the nuclear spin states in the $m_S = +1$ and $m_S = -1$ sublevels, in an external magnetic field oriented perpendicular to the hexagonal $c$-axis ($B \perp c$, $\theta = 90^\circ$). The in-plane orientation of the magnetic field is experimentally determined to be $\varphi = 12^\circ$ relative to the direction of one of the N(1) group nitrogen dangling bonds, which are interconnected by the $C_3$ rotational symmetry, as indicated by the dashed line in Fig.~1(a). Namely, the angle $\varphi$ is calculated using previously established tensors for the HF and NQ interactions of the $V_B^-$ spin with the atoms N(1)~\cite{ref24}, as shown in Fig.~S1 of the supplementary material~\cite{ref30}. 
\begin{figure}
    \centering
    \includegraphics[width=\linewidth]{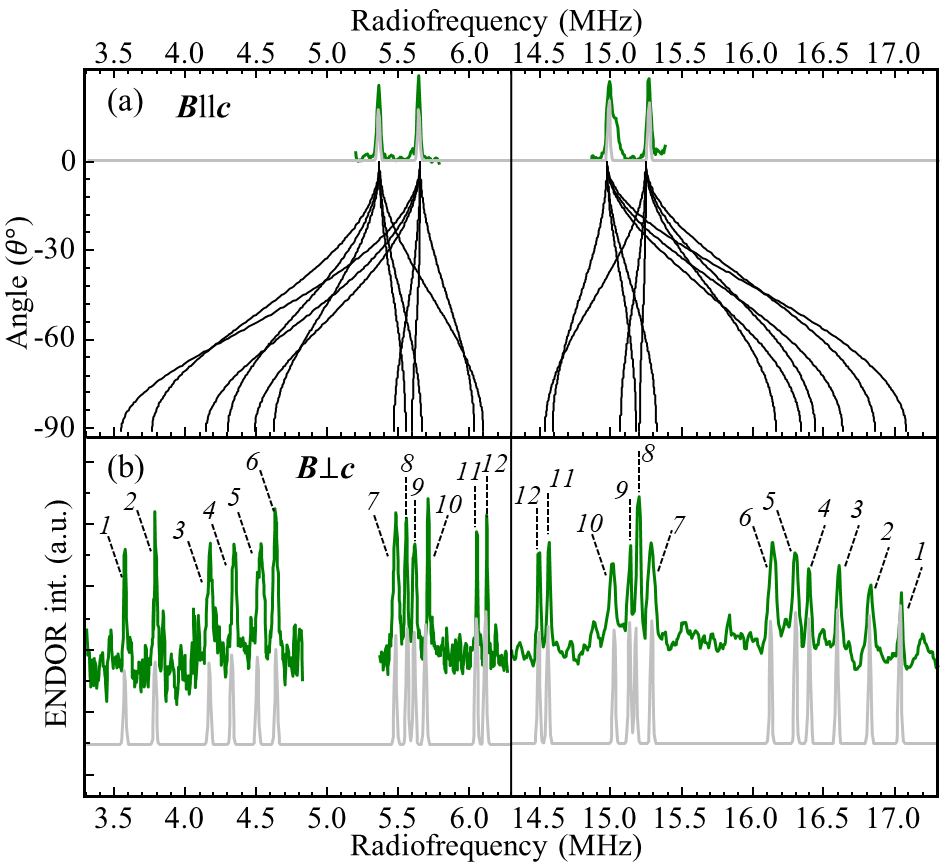} 
    \caption{(a) NMR transitions from Fig.~2 within the $m_S = +1$ and $m_S = -1$ spin states, measured at $B \parallel c$ orientation. Solid black lines show the calculated angular dependence of the NMR transitions as a function of the angle $\theta$ between the $c$-axis and the static magnetic field, as the field rotates away from $B \parallel c$ toward the ($0001$) plane, based on Eq.~(1). (b) Twelve NMR transitions labeled 1-12 within the $m_S = +1$ and $m_S = -1$ states, measured in $B \perp c$ orientation. Calculated spectra using Eq.~(1) are overlaid in grey and are made under the assumption that the magnetic field vector is $12^\circ$ away from the nitrogen dangling bond. Spin-Hamiltonian parameters used in the calculations are summarized in Table~I.}
    \label{fig:nmr_transitions}
\end{figure}

Additionally, we present the calculated evolution of the NMR frequencies as the magnetic field is rotated, varying the angle $\theta$ between the field $B$ and the $c$-axis from $B\parallel c$ ($\theta = 0^\circ$) to $B \perp c$, as described by Eq.~(1). These results are summarized in Fig.~3(a). The key observation from Fig.~3(b) is that each spectrum consists of six pairs of lines. Given that, in this magnetic field orientation ($B \perp c$, $\varphi = 12^\circ$), all in-plane nitrogen atoms are non-equivalent, we conclude that the spectra result from HF and NQ interactions with six distinct, non-equivalent nitrogen atoms. Indeed, each $^{14}$N nucleus contributes two NMR frequencies in the $m_S = +1$ or $m_S = -1$ states, as schematically shown in Fig.~1(d). This rules out the three nitrogen atoms in the N(2) shell, leading to the conclusion that the probed nuclear spins are located in the N(3) shell. Fitting the spectra using Eq.~(1) with the provided angle evolution of the spectra from $B \parallel c$ toward the $B \perp c$ orientation allows us to establish all components of the interactions as well as directions of the principal axis of the corresponding tensors, as indicated in Table~I. The direction of the principal $z$-axis of the HF and NQ tensor was chosen in the ($0001$) plane of the crystal, such that for each N(3) nucleus, the axis was tilted by $14^\circ$ from the direction of the N(1) center's $z$-axis in the perpendicular orientation, with an accuracy of up to $\pi$.
To validate the interpretation of the ENDOR data and explicitly demonstrate that our experiment probes the distant in-plane N(3) nuclear spins, we present the results of DFT calculations. The calculations were carried out for the $V_B^-$ placed in a $10 \times 10$ hBN monolayer (200 atoms) using the GIPAW module of the Quantum ESPRESSO software~\cite{ref31, ref32}, PBE exchange-correlation functional~\cite{ref33}, norm-conserving pseudopotentials~\cite{ref34}, and a plane-wave basis set with 600 eV kinetic energy cutoff. The results summarized in Fig.~4 show that the hyperfine couplings do not merely depend on the distance from the vacancy but are determined by the spin density distribution. 
\begin{figure}
    \centering
    \includegraphics[width=\linewidth]{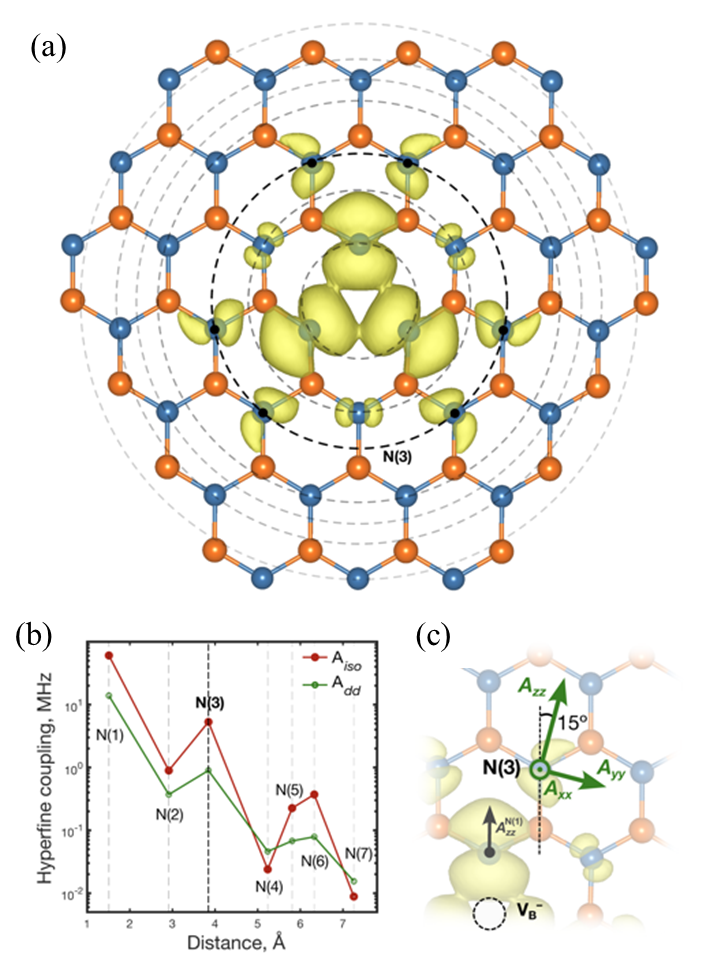} 
    \caption{(a) DFT calculated electron spin density distribution of the $V_B^-$ center in the hBN monolayer. (b) Isotropic, $A_{\text{iso}}$, and dipolar, $A_{\text{dd}}$, HF coupling constants calculated for different nitrogen shells and plotted against the distance from the center of the vacancy. The N(3), N(4), N(6) provide six nitrogen nuclei in the shell; the others are threefold. (c) Principal directions of the HF tensor of a representative N(3) atomic site (green arrows) referenced to the principal $zz$ direction of the nearest N(1) nucleus.}
    \label{fig:DFT_calculations}
\end{figure}
As demonstrated in Fig.~4a, the spin density distribution is, in turn, defined by the atomic structure and chemical bonding pattern. As a result, both the isotropic and dipolar HF coupling are stronger for the third nitrogen shell, N(3), than for the second one (cf. Fig.~4b). Starting from the fourth shell, the HF couplings further decrease compared to N(3). This observation is in line with the HF values reported in ~\cite{ref25}.  The HF and NQ coupling constants calculated for N(3) and listed in Table~I are found to be in very good agreement with the experimentally determined values, thus confirming the identification of the ENDOR results.
\begin{table}
\centering
\caption{HF and NQ parameters with $^{14}$N atoms in the N(3) shell (in MHz). $\eta$
is dimensionless. N(3) EXP, and N(3) DFT represent experimental and theoretical values, respectively. Isotropic and anisotropic parts of the HF are determined as follows: $A_{\text{iso}} = (A_{xx} + A_{yy} + A_{zz})/3$, $A_{dd} = (2A_{zz} - A_{xx} - A_{yy})/6$. For the resultung values for the first seven shells, see Fig. 4b.}
\begin{tabular}{p{2.5cm} p{2.5cm} p{2.5cm}}
\hline\hline
\textbf{Parameter} & \textbf{N(3) EXP.} & \textbf{N(3) DFT} \\
\hline
$A_{xx}$ & $4.41 \pm 0.02$ & 4.407 \\
$A_{yy}$ & $4.43 \pm 0.02$ & 4.409 \\
$A_{zz}$ & $6.86 \pm 0.02$ & 7.095 \\
$A_{\text{iso}}$ & $5.23 \pm 0.03$ & 5.304 \\
$A_{dd}$ & $0.82 \pm 0.03$ & 0.896 \\
$C_q$ & $0.37 \pm 0.03$ & 0.426 \\
$\eta$ & $0.55 \pm 0.01$ & 0.469 \\
\hline\hline
\end{tabular}
\end{table}
Finally, in agreement with the experiment, the $zz$ principal direction of the HF tensor of N(3) is perpendicular to the $c$-axis and tilted by $15^\circ$ from that of the N(1) nucleus (Fig.~4c). 

In this Letter, we demonstrate the first steps toward studying and reading out remote nuclear spins in hBN with the central $V_B^-$ $S$ = 1 electron spin. Building on recent advances in utilizing the $V_B^-$ defect as a quantum sensor \cite{ref16, ref17, ref18, ref19, ref20}, our demonstration broadens the application of this defect to sensing distant nuclear spins. We emphasize that understanding and uncovering the nuclear spin system in hBN requires relatively high magnetic fields and microwave frequencies. These conditions are essential for isolating and accurately identifying different nuclear spin groups, leveraging nuclear Zeeman interactions to achieve spectral separation. This approach lays the groundwork for coherent control and potential selective population transfer from the $V_B^-$ electron spin system to nuclear spin rings associated with nitrogen and boron isotopes surrounding the vacancy. The latter is key to implementing hBN's nuclear sublattices for future quantum simulators. \cite{ref21, ref22, ref23, ref35}.

The authors thank T. Biktagirov for fruitful discussions and assistance with DFT calculations. Financial support of the Russian Science Foundation under Grant RSF~24-12-00151 is acknowledged.

\bibliography{references}  

\end{document}